\documentclass[aps,pre,twocolumn,superscriptaddress,nofootinbib,showpacs,10pt]{revtex4-1}

\usepackage{amsmath}
\usepackage{amssymb}

\usepackage{graphicx}
\usepackage{epsfig}
\usepackage[utf8]{inputenc}
\usepackage{bm}

\begin{document}
\title{Thermal Radiation in Rayleigh-B\'{e}nard Convection Experiments
}
\author{P. Urban} 
\email{urban@isibrno.cz}
\author{T. Kr\'{a}l\'{i}k}
\author{P. Hanzelka}
\author{V. Musilov\'{a}}
\author{T. V\v{e}\v{z}n\'{i}k}
\affiliation{The Czech Academy of Sciences, Institute of Scientific Instruments, Kr\'{a}lovopolsk\'{a} 147, Brno, Czech Republic}
\author{D. Schmoranzer} 
\author{L.~Skrbek}
\email{skrbek@fzu.cz}
\affiliation{Faculty of Mathematics and Physics, Charles University, Ke Karlovu 3, Prague, Czech Republic}

\date{\today}

\begin{abstract}
An important question in turbulent Rayleigh-B\'{e}nard convection (RBC) is the effectiveness of convective heat transport, which is conveniently described via the scaling of the Nusselt number (${\rm{Nu}}$) with the Rayleigh (${\rm{Ra}}$) and Prandtl (${\rm{Pr}}$) numbers. In RBC experiments, the heat supplied to the bottom plate is also partly transferred by thermal radiation. This heat transport channel, acting in parallel with the convective and conductive heat transport channels, is usually considered insignificant and thus neglected. Here we present a detailed analysis of conventional far-field as well as strongly enhanced near-field radiative heat transport occurring in various RBC experiments, and show that the radiative heat transfer partly explains differences in ${\rm{Nu}}$ measured in different experiments. A careful inclusion of the radiative transport appreciably changes the ${\rm{Nu}}={\rm{Nu}}({\rm{Ra}})$ scaling inferred in turbulent RBC experiments near ambient temperature utilizing gaseous nitrogen and sulphur hexafluoride as working fluids. On the other hand, neither the conventional far-field radiation nor the strongly enhanced near-field radiative heat transport appreciably affects the heat transport law deduced in cryogenic helium RBC experiments.
\end{abstract}
\maketitle

\section{Introduction}
The very useful model for fundamental studies of buoyancy driven flows --- Rayleigh-B\'{e}nard convection (RBC) --- occurs in a fluid layer confined between two laterally infinite, perfectly conducting plates heated from below in a gravitational field; for review, see Refs.~\cite{Rev,CHS}. For an Oberbeck-Boussinesq fluid of constant physical properties except the density that linearly depends on temperature, it is fully characterized by the Rayleigh number, ${\rm{Ra}}=g \alpha\Delta T L^3 /(\nu\kappa)$, and the Prandtl number, ${\rm{Pr}}=\nu/\kappa$. Here $g$ stands for the acceleration due to gravity, and $\Delta T= T_B-T_T$ is the temperature difference between the parallel bottom and top plates separated by the vertical distance $L$. The working fluid is characterized by the thermal conductivity, $\lambda$, and by the combination $\alpha/(\nu \kappa)=\eta$, where $\alpha$  is the isobaric thermal expansion, $\nu$  is the kinematic viscosity, and $\kappa$ the thermal diffusivity. Experimental investigations of RBC of lateral dimension $D$ involve an additional important parameter, the aspect ratio $\Gamma=D/L$.  The key feature of RBC is its ability to transfer heat from the heated bottom plate to the cooled top plate. The convective heat transfer effectiveness is conventionally described using the Nusselt number, ${\rm{Nu}} = Q_B/Q_M=L \dot q/(\lambda \Delta T)$, via the ${\rm{Nu}} = {\rm{Nu}}({\rm{Ra}}; {\rm{Pr}}; \Gamma)$ scaling \citep{NS2003,Rev,CHS}, where $\lambda$ denotes the thermal conductivity of the working fluid. The Nusselt number is simply the ratio of the convective heat transport, $Q_B$, accomplished by a flowing working fluid, such as air, water, ethane, SF$_6$, glycerin, or helium, in comparison with the amount possible solely due to their molecular conduction, $Q_M$.

In RBC experiments, extreme care must be taken to correctly evaluate ${\rm{Nu}}$ and ${\rm{Ra}}$ in order to account for the influence of various factors \citep{NS2003}, such as parasitic heat leaks, finite thermal conductivity and capacity of plates \citep{Versicco2004} and walls \citep{Roche2001} of the RBC cell, non Oberbeck-Boussinesq effects, adiabatic gradient in the working fluids as well as uncertainties in their physical properties, especially in the vicinity of the equilibrium saturation curve \citep{OurPRE} and the critical point \citep{OurNJP} of the working fluid. This is usually done via corrections that follow various models of heat flow in a particular RBC cell.

In this work, we turn attention to the well-known fact that heat supplied to the bottom plate is also partly transferred by thermal radiation. Consequently, there is another heat transport channel acting in parallel with the convective and conductive heat transport channels. This heat flow channel, extremely important when considering thermal convection in Sun or stars, is not taken into account or considered inefficient in most laboratory RBC experiments \citep{Hogg}, and thus neglected. Here we show that in many RBC experiments, radiative heat transfer ought to be taken into account, as it partially explains the differences in measured ${\rm{Nu}}$. Thus, it changes appreciably the exponent $\gamma$ in the scaling relation ${\rm{Nu}} \propto {\rm{Ra}}^\gamma$ observed in ambient temperature turbulent RBC experiments utilizing gaseous helium, nitrogen or sulphur hexafluoride as working fluids. Furthermore, we consider the strongly enhanced near-field radiative heat transport at cryogenic temperatures which, in principle, might affect, e.g., the heat transport through RBC cells with closely spaced plates designed to detect the onset of convection~\cite{LeesPRL,Lees,Metcalfe} and show that, in cryogenic helium RBC experiments, neither the conventional far-field nor the strongly enhanced near-field radiative heat transport appreciably affects the total heat transfer.

\section{Radiative heat transfer}
Any surface of a body at temperature $T$ radiates in all directions comprising a hemisphere the total heat power per unit area $e \sigma T^4$, where $e$ is the total hemispherical emissivity and $\sigma = 5.67 \times 10^{-8}$ Wm$^{-2}$K$^{-4}$ denotes the Stefan-Boltzmann constant. The emissivity $e$ varies between 0 (perfect reflector) and 1 (perfect blackbody). At the same time, such a surface reflects heat intensity $(1-e)H$, where $H$ is the intensity of incident radiation per unit area.

Basic laws of Planck and Stefan-Boltzmann for thermal radiation assume that the radiative heat exchange occurs over distances larger than the relevant wavelengths of the thermal radiation. The radiative heat transfer between laterally infinite plane parallel surfaces in vacuo is then independent on the distance $L$ between them; this conventional case will further be quoted as far field radiation (FF). However, it was shown experimentally \cite{Kralik2012} that with decreasing $L$ the radiative heat exchange between conductive surfaces increases and can exceed the FF heat flux by more than two orders of magnitude. The reason is that, in parallel with the thermal radiation, there are evanescent waves propagating along the surface of the material (plates of the RBC cell), whose energy exponentially decreases with the distance from the surface. When the field of evanescent waves reaches the opposite surface (opposite plate of the RBC cell), the energy can be transferred by tunneling of photons. We shall quote this peculiar radiative heat transfer as near field (NF) radiation. The distance $L^*$ below which the NF becomes dominant can be estimated \citep{Polder} as $L^*\cong \lambda_{BB}= c_0 \hbar/(k_B T)$, where the $\lambda_{BB}$ is characteristic wavelength of the blackbody radiation at temperature $T$, $c_0$ denotes the speed of light in vacuum, $\hbar$ and $k_B$ are respectively the reduced Planck and Boltzmann constants. While at room temperature $\lambda_{BB}$ is of order 10~$\mu$m, suggesting that in typical RBC experiments performed at ambient temperatures the NF radiative heat transfer is insignificant, for RBC experiments performed at cryogenic temperatures (i.e., at about 2-10 K) $\lambda_{BB}$ becomes of order of 1 mm, comparable with the size of some experimentally used RBC cells and the NF radiative heat transfer might in principle become relevant. We shall discuss this issue, based on our own dedicated measurements of NF radiative heat exchange between closely spaced copper surfaces \citep{Veznik}.

\subsection{FF Radiative heat transfer in RBC experiments}

Most experimental studies of turbulent RBC have been performed in cells of aspect ratios about unity, of sizes ranging typically from 2~cm to 2~m, i.e., much larger than $\lambda_{BB}$. Consequently, the NF radiative transfer is hardly relevant. Experimental RBC cells are filled with various liquids or gases as working fluids. In the following we focus on gaseous media only, as liquids typically exhibit relatively large thermal conduction. Therefore, radiative heat transfer in liquids will generally not be significant, unless the liquid is radiatively participating.

Furthermore, our model (see the Appendix) of FF radiative heat transfer in RBC experiments performed with gaseous working fluids around room temperature assumes that these gases are fully transparent at the relevant wavelengths, i.e., around 10~$\mu$m according to the Wien displacement law. This is justified for helium, nitrogen, and argon, but not necessarily for CO$_2$ and SF$_6$. These belong among the so-called greenhouse gases, SF$_6$ being the most potent one, and have several absorption lines in the infrared spectral range \citep{Rothman}. Such gases will be discussed separately, as absorption of radiation results in volumetric heating of the working fluid, facilitating a heat flux that bypasses the thermal boundary layer, which is crucial in the present understanding of RBC. To this effect, we note that external FF radiative heating combined with volumetric absorption was recently exploited \cite{Lepot} in order to avoid steep temperature gradients occurring near the plates, leading to the ultimate ${\rm{Nu}} \propto {\rm{Ra}}^{1/2}$ scaling; however, the full analysis of this class of experiments is beyond the scope of this study.

To calculate the FF radiative part of heat transfer in room temperature RBC experiments performed in cylindrical cells of aspect ratio $\Gamma$ with transparent working fluids, we utilize the model proposed by Hogg \cite{Hogg}. In order to calculate the correction to the total heat transfer through the RBC cell due to FF radiation, one needs to determine the value of incident radiation per unit area, $H$, incoming to the bottom plate originating from other surfaces of the cell, i.e., from the top plate and the cylindrical wall. This depends on the temperature and emissivity of these surfaces and on the aspect ratio $\Gamma$ of the particular cell and can be calculated in a standard way, using the so-called view factors $F$ for cylindrical geometry \citep{Siegel}. For clarity of presentation, the derivation of the model is given in the Appendix. The model assumes top and bottom plates of temperatures $T_T$ and $T_B$, of emissivity $e_T$ and $e_B$ and a fully transparent working fluid of temperature $T_M=1/2(T_T+T_B)$ everywhere in the cell, in tight thermal contact with the cylindrical wall of the same temperature $T_M$ and emissivity $e_W$. This simple approximation is justified for turbulent high ${\rm{Ra}}$ RBC experiments, for which the mean temperature drops occur over thin boundary layers adjacent to plates while the bulk of the cell has (nearly) the same temperature $T_M$.

\begin{table*}
	\begin{tabular}{c|c|c|c|c|c|c|c|c|c}
    &  $D$  & $L$  &  $T_M$  &  $\Delta T $  &  $e_B$  & $Q_B$  & $Q_1$  &  $Q_B^{corr}$  &  {\rm{Ra}}\\
    &  (m)  & (m)  &   (K)   &   (K)         &         & (W)    & (W)    &  \%          &     \\
    & & & & & & & & & \\
    Brno: $\Gamma= 1$ & 0.30 & 0.30 & 4.9 & 0.24 & 0.05 & 0.87 & $1.2 \times 10^{-8}$ & $1.3 \times 10^{-6}$ & $1.6 \times 10^{13}$\\
    \emph{cryogenic $^4$He gas} & 0.30 & 0.30 & 4.9 & 0.03 & 0.05 & 0.06 & $1.9 \times 10^{-9}$ & $2.5 \times 10^{-6}$ & $2.2 \times 10^{12}$\\
	& & & & & & & & & \\
    Santa Barbara: $\Gamma= 0.5$ & 0.25 & 0.49 & 298 & 20.1 & 0.05 & 34.6 & 0.16 & 0.5 & $6.6 \times 10^{9}$\\
    \emph{room temperature $^4$He gas} & 0.25 & 0.49 & 298 & 5.0 & 0.05 & 2.0 & 0.04 & 1.9 & $4.3 \times 10^{7}$\\
    & & & & & & & & & \\
    G\"{o}ttingen: $\Gamma= 1$ & 1.12 & 1.12 & 295 & 16.5 & 0.05 & 596.7 & 2.48 & 0.4 & $1.4 \times 10^{14}$\\
    \emph{room temperature SF$_6$ gas} & 1.12 & 1.12 & 295 & 3.0 & 0.02 & 13.8 & 0.17 & 1.3 & $3.7 \times 10^{11}$\\
                               & 1.12 & 1.12 & 295 & 3.0 & 0.05 & 13.8 & 0.44 & 3.2 & $3.7 \times 10^{11}$\\
                               & 1.12 & 1.12 & 295 & 3.0 & 0.08 & 13.8 & 0.70 & 5.1 & $3.7 \times 10^{11}$\\
    & & & & & & & & & \\
    G\"{o}ttingen: $\Gamma= 0.5$ & 1.12 & 2.24 & 295 & 13.6 & 0.05 & 523.9 & 2.02 & 0.4 & $1.1 \times 10^{15}$\\
    \emph{room temperature SF$_6$ gas} & 1.12 & 2.24 & 297 & 1.0 & 0.02 & 3.9 & 0.06 & 1.5 & $1.5 \times 10^{12}$\\
                                 & 1.12 & 2.24 & 297 & 1.0 & 0.05 & 3.9 & 0.15 & 3.9 & $1.5 \times 10^{12}$\\
                                 & 1.12 & 2.24 & 297 & 1.0 & 0.08 & 6.2 & 0.24 & 3.9 & $1.5 \times 10^{12}$\\
\end{tabular}
\caption{Examples of the role of FF radiative heat transfer in selected RBC experiments performed in cylindrical cells of various sizes (diameter $D$, height $L$ and aspect ratio $\Gamma=D/L$). Real data points, corresponding to various selected $\Delta T$, $T_M$, applied heat to the bottom plate $Q_B$ and ${\rm{Ra}}$ are considered, with realistic emissivity values $e_B=e_T$ and $e_W=0.9$. The corresponding FF net radiative heat flux is given in the column $Q_1$, and its relative weight in column $Q_B^{corr}$. We see that the FF radiative correction varies from being nearly negligible (especially in comparison with other corrections such as that due to walls) to being several percent of the total, appreciably affecting ${\rm{Nu}}$ as well as local ${\rm{Nu}} \propto {\rm{Ra}}^\gamma$ scaling exponent $\gamma$, as shown graphically in Figs.~\ref{FigRoomHalf} and~\ref{FigRoomOne}.}
\label{tab:examples}
\end{table*}

\begin{figure}[th]
\centering
\includegraphics[width=0.99\linewidth]{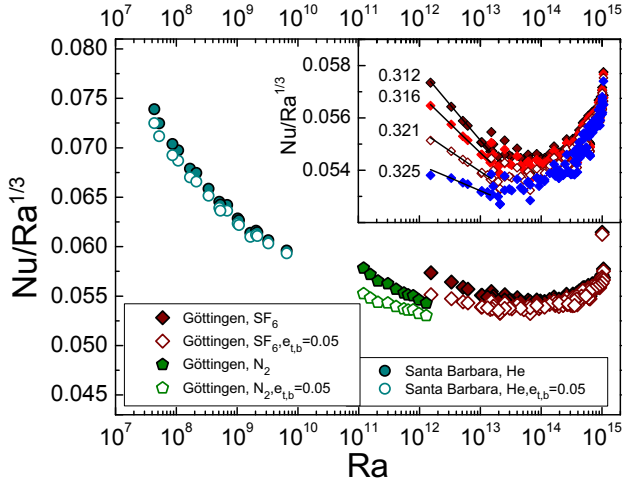}
\caption{Main panel: The compensated values ${\rm{Nu}}\, {\rm{Ra}}^{-1/3}$ plotted versus ${\rm{Ra}}$ obtained at ambient temperatures in cylindrical aspect ratio $\Gamma=1/2$ cells, displayed without (filled symbols) and with corrections to FF radiative heat transfer, assuming realistic emissivity values $e_B=e_T=0.05$  taken from Refs.~\cite{Ablewski,Hawks,BrnoDatabase} for the top and bottom copper or aluminium plates and $e_W=0.9$ for the plexiglass wall (open symbols), evaluated using the model introduced by Hogg~\cite{Hogg}; for details, see the Appendix. The model assumes a fully transparent working fluid, which is the case of the Santa Barbara He data \cite{Hogg} and G\"{o}ttingen N$_2$ data \cite{AhlersSearch}; the G\"{o}ttingen SF$_6$ data in Ref.~\cite{AhlersNJPhalf} are affected both by FF radiative heat transfer and by partial absorption of radiation by SF$_6$. See the text for further details. Inset: From top to bottom, the original G\"{o}ttingen SF$_6$ data of Ref.~\cite{AhlersNJPhalf} and the same data corrected for FF radiative heat transfer using various values of $e_B=e_T= 0.02; 0.05;$ and $0.08$, illustrating the influence of the FF correction on the values of scaling exponents for local {\rm{Nu}}({\rm{Ra}}), evaluated for ${\rm{Ra}} < 2 \times 10^{13}$.}
\label{FigRoomHalf}
\end{figure}

\begin{figure}[th]
\centering
\includegraphics[width=.99\linewidth]{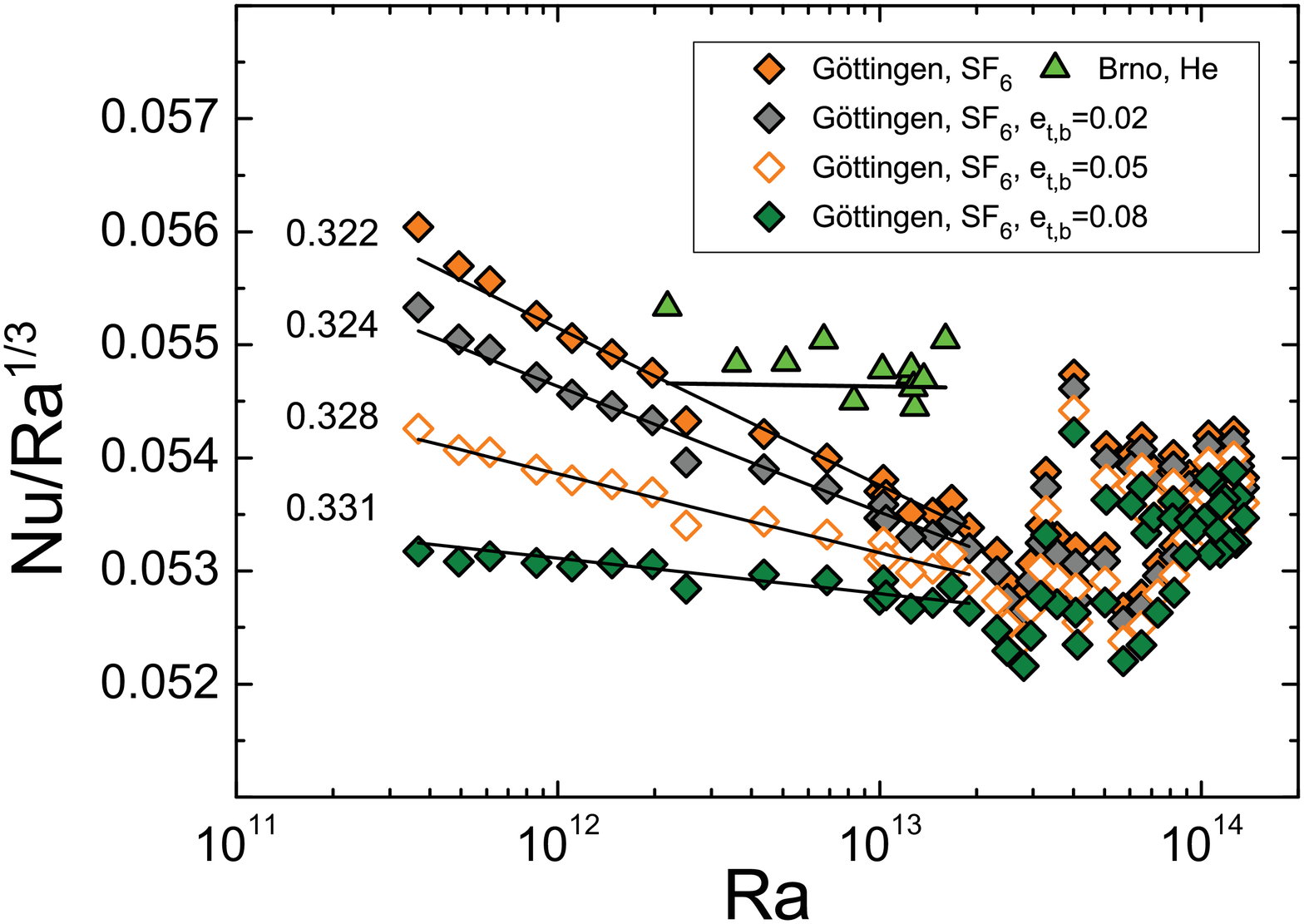}
\caption{The compensated values ${\rm{Nu}}\, {\rm{Ra}}^{-1/3}$ plotted versus ${\rm{Ra}}$ for the data obtained in cylindrical aspect ratio $\Gamma=1$ cells at cryogenic conditions \cite{OurPRE} (green triangles) and at ambient temperatures \cite{AhlersG1}. The latter are shown, from top to bottom series, without and with corrections to FF radiative heat transfer, assuming the copper plates emissivity $e_T=e_B= 0.02; 0.05;$ and $0.08$  taken from Refs.~\cite{Ablewski,Hawks,BrnoDatabase} and the plexiglass wall emissivity $e_W=0.9$. The corrections shown assume a fully transparent working fluid. Note the influence of corrections to FF radiative heat transfer on the shown values of local {\rm{Nu}}({\rm{Ra}}) scaling exponents evaluated for ${\rm{Ra}} < 2 \times 10^{13}$.}
\label{FigRoomOne}
\end{figure}

\subsection{Corrections due to radiative FF heat transfer in selected high $\rm{{\rm{Ra}}}$ RBC experiments}
Let us emphasize that for high ${\rm{Ra}}$ RBC experiments utilizing cryogenic helium gas and performed at temperatures of a few kelvin \cite{Wu,ChavannePRL,Chavanne2001,Roche2010,Nature,Wind,NS2003,Niemela2010,BrnoPRL1,BrnoPRL2,OurNJP,OurPRE} the FF radiative heat transfer is completely negligible, thanks to its strong temperature dependence ($\propto T^4$). This is confirmed by two examples of corrections calculated for our own $\Gamma= 1$ cell 30 cm in diameter used in most Brno RBC experiments, see Tab.~\ref{tab:examples}.

In order to appreciate the influence of radiative FF heat transfer in room temperature high ${\rm{Ra}}$ RBC experiments quantitatively, we have selected experiments performed in cylindrical cells of aspect ratio $\Gamma=1/2$ and $\Gamma=1$. The results of our analysis are illustrated in Figs.~\ref{FigRoomHalf} and \ref{FigRoomOne}; for convenience some typical values are given in Tab.~\ref{tab:examples}. The corrections shown are evaluated as follows: using the FF radiation approach of Ref.~\cite{Hogg} (see the Appendix), for each data point we calculate the radiative part of the heat flux and subtract it from the total supplied heat flux experimentally applied to the bottom plate, in order to separate the FF radiative heat transport from that carried by conduction and convection, $Q_b$, mediated by the working fluid, and evaluate the corrected ${\rm{Nu}}$.

Let us start with the $\Gamma=1/2$ experiments, see Fig.~\ref{FigRoomHalf}. We first confirm the conclusion of Ref. \cite{Hogg} that for the particular case of the Santa Barbara helium data ranging from $4.34 \times 10^{7}$ to $6.55 \times 10^{9}$ in ${\rm{Ra}}$, the corrections are indeed relatively unimportant, although they slightly change both the values of $\rm{{\rm{Nu}}}$ and the local scaling exponent $\gamma$ introduced above. On the other hand, the corrections due to radiative FF heat transfer would be more important in high $\rm{{\rm{Ra}}}$ RBC experiments performed in much larger RBC cells placed in the so-called U-boat of G\"{o}ttingen. The main panel of Fig.~\ref{FigRoomHalf} shows the published data together with the corrected ones, assuming fully transparent working fluid and realistic emissivity values $e_b=e_t=0.05$ \cite{Ablewski,Hawks,BrnoDatabase} for the top and bottom copper plates and $e_w=0.9$ for the plexiglass wall (open symbols). These corrections appreciably alter both the values of ${\rm{Nu}}$ and the local scaling exponent.

Let us focus on the low end of the data range in Ref.~\cite{AhlersNJPhalf}. The correction is significant and, moreover, appreciably changes the ${\rm{Nu}} \propto {\rm{Ra}}^\gamma$ scaling. Indeed, if we evaluate the local scaling exponent $\gamma$ over the same range of {\rm{Ra}}  $(1.53 \times 10^{12} \leq {\rm{Ra}} \leq 1.45 \times 10^{13})$, using the same data points of Ref.~\cite{AhlersNJPhalf}, for values of the copper or aluminium plates emissivity $e_T=e_B=0$; 0.02; 0.05; 0.08 and the plexiglas wall emissivity $e_W=0.9$, the scaling exponent increases from 0.312 to values shown in the inset of Fig.~\ref{FigRoomHalf}. We see that in this particular case the realistic correction for the local scaling exponent $\gamma$ to the FF heat transfer is larger than the last significant digit of $\gamma$ quoted in Ref.~\cite{AhlersNJPhalf}.

For higher ${\rm{Ra}}$ range covered by the same data of Ref.~\cite{AhlersNJPhalf}, the correction due to the FF heat transfer becomes less significant, thanks to considerably larger heat flux supplied to the bottom plate. Moreover, it was shown recently \cite{OurPRE} that the data belonging to the upper end of the range of ${\rm{Ra}}$ covered by Ref.~\cite{AhlersNJPhalf} are most likely affected by the choice of the working point in the p-T phase diagram of SF$_6$: together with the imposed temperature drop between plates, the top plate temperature $T_T$, at the pressure used in the experiment, is too close to the equilibrium saturated vapour curve of SF$_6$. We therefore refrain from commenting here on the scaling exponent for the high  ${\rm{Ra}}\gtrsim 10^{14}$ data obtained in this experiment.

Let us now discuss the selected $\Gamma=1$ experiments (see Fig.~\ref{FigRoomOne}), comparing our own cryogenic helium data \citep{OurPRE} with the G\"{o}ttingen SF$_6$ data in Ref.~\cite{AhlersG1}. For the same reason as above, we refrain from commenting these for ${\rm{Ra}}> 2 \times 10^{13}$.  It is important to emphasize, as argued in Ref.~\cite{OurPRE}, that these two sets of data have been measured using the same experimental protocol, obtained under nominally similar conditions, thus allowing to illustrate directly the importance of corrections due to radiative FF heat transfer in room temperature high ${\rm{Ra}}$ RBC experiments by comparing the data with the cryogenic experiment where the radiative heat transport is vanishingly small.

We have shown \citep{BrnoPRL1} that for $7.2 \times 10^6 \leq {\rm{Ra}} \leq 10^{11}$ our sidewall-corrected data agree with suitably corrected data from complementary cryogenic experiments, and are consistent with $ {\rm{{\rm{Nu}}}}\propto  {\rm{Ra}}^{2/7}$. On approaching $ {\rm{Ra}}\approx 10^{11}$, all cryogenic data display a broad crossover to $ {\rm{Nu}}\propto {\rm{Ra}}^{1/3}$, as predicted theoretically~\cite{Malkus,Priestley}, and in accord with the theoretical model of Ref.~\cite{GrLohse} and its update in Ref.~\cite{GrLohseB}. Fig.~\ref{FigRoomOne} indicates that the scaling exponent of exactly 1/3 (matching that of Refs. \cite{BrnoPRL1,OurPRE}) would be formally achieved in the SF$_6$ experiments \cite{AhlersG1} for $e_T=e_B \simeq 0.1$. Such a high emissivity is unlikely for high-quality surface finish of the G\"{o}ttingen copper and aluminium plates; the realistic value is probably 2-3 times lower \cite{Ablewski,Hawks,BrnoDatabase}. On the other hand, SF$_6$ is not fully transparent for thermal radiation, a property we shall consider next.

\begin{figure}[th]
\centering
\includegraphics[width=.99\linewidth]{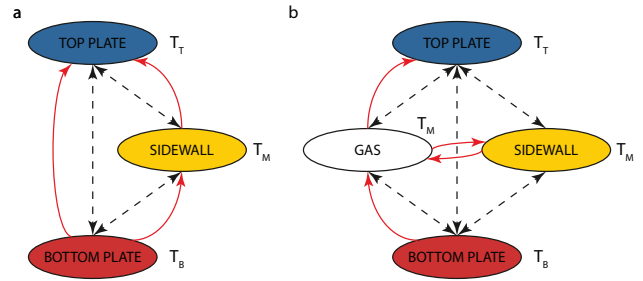}
\caption{Schematic representations of heat transfer for the case of transparent (a) and participating gas (b). Straight dashed black arrows represent radiative heat transfer, while curved solid red arrows represent convection. Radiative heat exchange between the gas and the sidewall is disregarded here, as they are assumed to have the same temperature $T_M$.}
\label{fig:thermalmodels}
\end{figure}

\subsection{Comments on absorption of thermal radiation by working fluids used in RBC experiments}
An RBC working fluid that absorbs, emits and scatters thermal radiation is called a participating medium \citep{Siegel}.
Unlike solids, gases do not have continuum infrared absorption or emission spectra, but interact with radiation only at selected wavelengths as determined by their molecular rotational and vibrational spectra. The complexity of such spectra is determined by the number of modes of the gas molecule and are fairly straightforward for monatomic or diatomic molecules, but reach a high degree of complexity for larger molecules such as SF$_6$. Hence, instead of seeking a full rigorous description of the interaction of SF$_6$ with thermal radiation, requiring calculations of detailed balance for every single rotational-vibrational mode using their respective Einstein coefficients, qualitative arguments will be given, supported by models of absorption of thermal radiation in SF$_6$ based on available absorption spectra in the HITRAN 2008 molecular spectroscopic
database\footnote{HITRAN database: https://hitran.org/}~\cite{Rothman} and in Ref.~\cite{Chapados}.

To begin with, let us compare the radiative heat transfer mechanisms in the zero absorption case, and in the case of significant absorption in the gas in the RBC cell. These two models are schematically shown in Fig.~\ref{fig:thermalmodels}, both consisting of the top plate, at temperature $T_{\rm T}$, bottom plate at $T_{\rm B}$, the cylindrical wall at $T_{\rm M}$. Additionally, in the absorbing case, the gas at $T_{\rm M}$ needs to be included. While the presence of the absorbing gas weakens direct radiative heat transfer between the plates and the walls, it opens an additional channel for the radiated energy to be transported to the top plate via interaction with the gas. We stress here that convective heat transfer occurs in parallel with radiative transfer and has significantly higher effectiveness, hence the energy of any radiation emitted by the bottom plate and absorbed by the gas outside the boundary layer is very likely to be transferred to the top plate (or walls) via convection rather than radiation.

In the zero absorption case, the net power transported from the bottom plate to the top plate via radiation is given solely by the temperatures, emissivities and view factors of the respective plates and the sidewall as discussed above. In steady state, the net radiative heat transfer is practically equal to the difference between the radiation power emitted by the top and bottom plates as the sidewall will maintain a stable temperature very close to $T_M$, emitting the same amount of energy per unit time as it absorbs. Focusing hence on the direct radiative heat transfer between the two plates, the situation is influenced mainly by the very low emissivities of the two plates ($e_T=e_B\simeq 0.05$). Hence, any radiation emitted by one plate in the direction of the other likely travels the span of the cell several times before being absorbed by either plate. On the practical level, this means that while the two plates contribute to the radiation field in the cell unequally (the hotter bottom plate emits significantly more radiation), the radiation absorption rates of both plates are \emph{nearly equal}, as the amount of incident radiation is very similar and the emissivities are the same.

Here we argue that this balance may be significantly altered by absorption of radiation inside the gas. For the time being, let us assume that a notable fraction of the emitted radiation is absorbed in the bulk gas outside the boundary layers. Energetically speaking, part of the absorbed radiation will be re-emitted, but the rest of it will increase the internal energy of the gas. This excess energy will then be carried around by the convective flow and will be preferentially transferred to the colder top plate. Hence we claim that significant absorption of radiation by the gas will lead to an \emph{increase} in the amount of energy transferred radiatively, necessitating a larger correction than that given by the model of Hogg~\cite{Hogg}.

\begin{figure}[th]
\centering
\includegraphics[width=.99\linewidth]{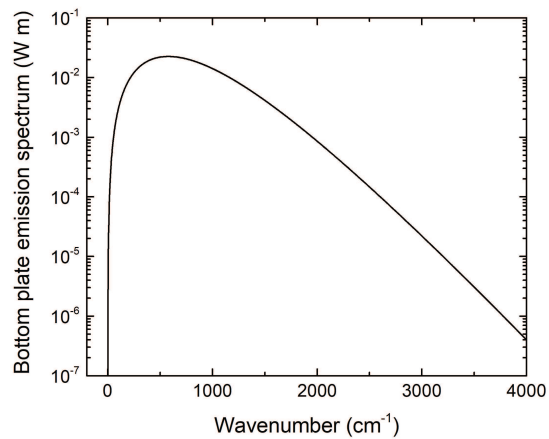}
\caption{Emission spectrum of the bottom plate of diameter 1.12~m at T = 295~K with emissivity $e=0.05$ calculated according to the Planck law. The total emitted power adds up to 21.2~W. Wavenumbers between 4000 and 6500~cm$^{-1}$ represent only $\approx$ 3\% of the emitted radiative power, and are not shown here (although they are included in the calculations).}
\label{fig:AbsSF6}
\end{figure}

\begin{figure}[th]
\centering
\includegraphics[width=.99\linewidth]{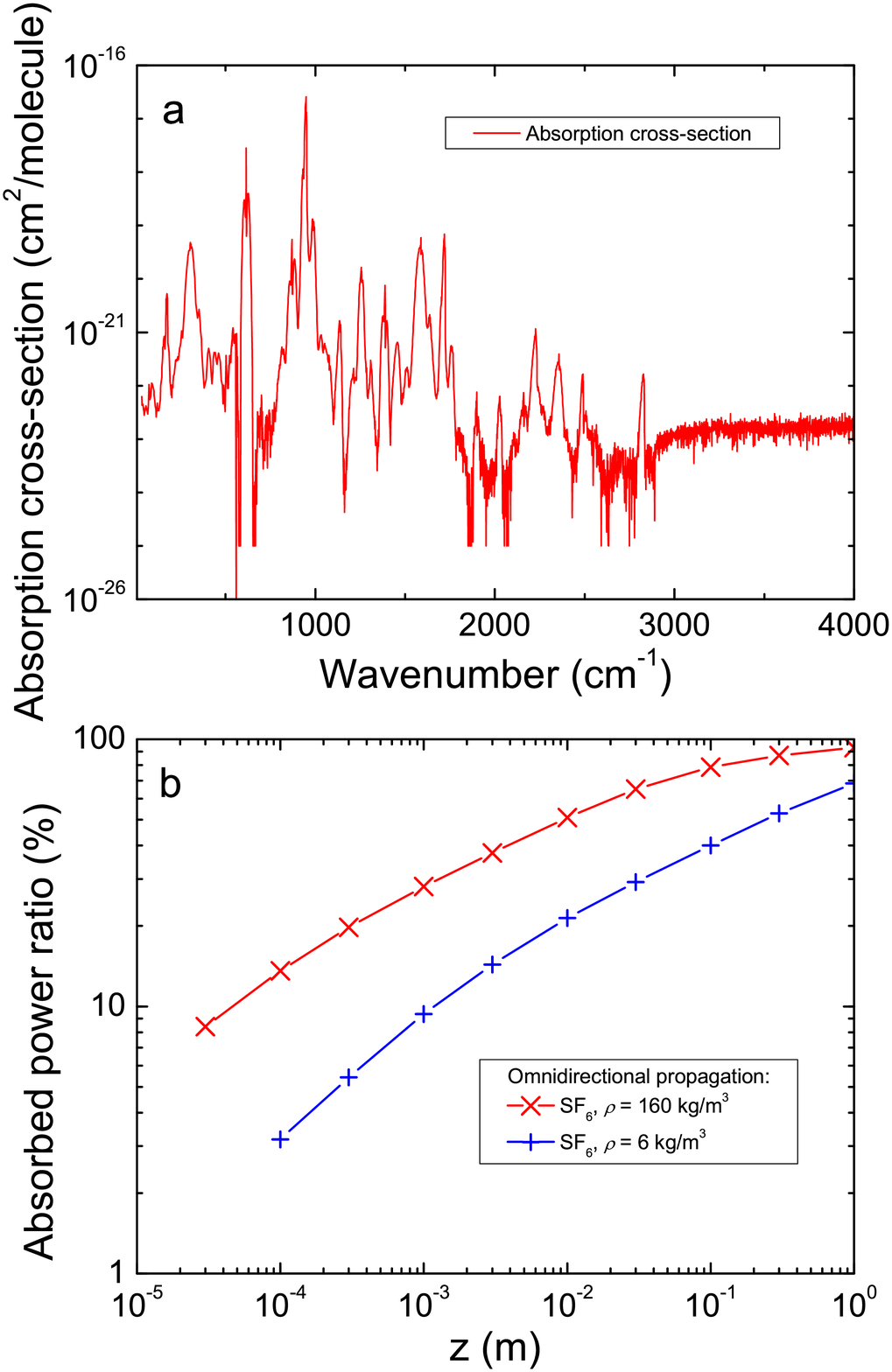}
\caption{a) Absorption cross-section of SF6 adapted from Ref.~\cite{Chapados} for wavenumbers from 32~cm$^{-1}$ to 570~cm$^{-1}$ and from the HITRAN database \cite{Rothman} for wavenumbers above 560~cm$^{-1}$. b) Absorbed radiative power ratio with respect to the heat power emitted by the bottom plate of diameter 1.12~m at T = 295~K (with constant emissivity across the relevant wavelengths -- ``gray surface''), plotted versus distance $z$ from the emitting plate. Absorption was calculated for two values of density of SF$_6$ (6~kg m$^{-3}$ and 160~kg m$^{-3}$), corresponding to atmospheric pressure and elevated pressure of 20~bar realised in the G\"{o}ttingen experiments. Distances of the order 1~mm and less estimate the boundary layer thickness, while the distance of 1~m corresponds to the height of the RBC cell with aspect ratio $\Gamma$ = 1. For both atmospheric and elevated pressure, most of radiation power is absorbed in the gas outside the boundary layer. Thermal radiation by the SF$_6$ gas itself was omitted in this calculation.}
\label{fig:AbsSF6A}
\end{figure}

To quantify the absorption of thermal radiation by SF$_6$, let us consider the case of omnidirectional thermal radiation emitted from a plane at the temperature of 295~K. The HITRAN database \cite{Rothman} and Ref.~\citep{Chapados} contain absorption spectra of SF$_6$ for wavenumbers between 32~cm$^{-1}$ and 6500~cm$^{-1}$. This practically covers the entire significant spectrum of thermal radiation emitted by a body at 295~K. In Figs.~\ref{fig:AbsSF6} and \ref{fig:AbsSF6A}a, we show the emission spectrum of the bottom plate as well as the absorption cross-section of SF$_6$.

We calculate the absorbed energy at distances up to 1~m from the emitting plate and show the results in Fig.~\ref{fig:AbsSF6A}. It is clearly confirmed that absorption of thermal radiation from a body at 295~K in SF$_6$ is indeed significant and that the above scenario holds. This is consistent with the reported observation that thermal radiation can accelerate the onset of the ultimate regime of convection~\cite{Lepot} by means of helping to bypass the boundary layers.

We note that the role of a thin layer of participating medium (CO$_2$) has also been investigated~\cite{Hutchison} at the onset of convection (up to ${\rm{Ra}} = 10^4$), where a shift of the convection onset and a suppression of heat transfer compared to a nearly transparent medium (air) has been observed. The present situation, however, differs in two important aspects: (i) the existence of a turbulent convection zone separated from the plates by thermal boundary layers, and (ii) the aspect ratio of the cell $\Gamma = 1/2$. The overall situation is clearly more complex and warrants dedicated work, both theoretical and experimental.

\subsection{NF radiative heat transfer in cryogenic RBC experiments}

\begin{figure}[th]
\centering
\includegraphics[width=.9\linewidth]{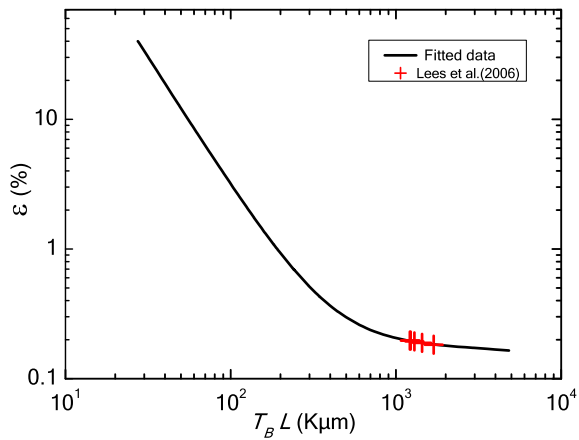}
\caption{Summary of measurements of the coefficient, $\varepsilon$, of the NF radiative heat transfer between closely spaced copper surfaces obtained from a series of our dedicated cryogenic experiments \citep{Veznik}. All our data resulting from two different ways of measurement (performed with constant temperature of the cold disc $T_1$, varying the distance $L$ or the temperature $T_2$ of the hotter disc) collapse when $\varepsilon$ is plotted versus the product $L T_B$, where $T_B$ effectively plays the role of the hotter bottom plate of the flat RBC cell. This allows to determine the working points for relevant cryogenic large aspect ratio RBC cells and estimate the role of NF radiation in the total heat transport. As an example, the red crosses mark these working points for the experiments described in Ref.~\cite{Lees}. For details, see the text.}
\label{NierFieldCu}
\end{figure}

We have already argued that the corrections due to the ``standard" FF radiative heat transfer in typical high ${\rm{Ra}}$ cryogenic RBC experiments are negligibly small. On the other hand, cryogenic RBC experiments aiming to determine the onset of convection have been performed in flat high aspect ratio cells, where strong enhancement of the radiative heat transfer could take place. At the same time, quoting Metcalfe~\cite{Metcalfe}: ``The critical Rayleigh numbers for many of the cryogenic experiments are found to differ from the standard prediction of 1708 by amounts which appear to be in excess of experimental error."
In order to shed some new light on this long-standing puzzle, we have performed dedicated cryogenic experiments. In short, we prepared a pair of circular sapphire discs 2.7 mm thick 35 mm in diameter, with their facing surfaces covered by sputtered Cu films 720 nm in thickness (the Cu film cannot be too thin, as it would be transparent for long wavelengths of the thermal radiation). We have utilized our cryogenic apparatus for study of near-field heat transfer \cite{Kralik2011} and measured the coefficient $\varepsilon$ of the radiative heat transfer for these two closely spaced flat circular Cu surfaces, defined as
\begin{equation}
\label{coeff}
\varepsilon = \frac{Q_{exp}} {\sigma S (T_2^4-T_1^4)}\,,
\end{equation}

\noindent
where $Q_{exp}$ is the directly measured heat flux and $S$ denotes the area of the disc. We found \citep{Veznik} that all our data resulting from two different ways of measurements (performed with constant temperature of the cold disc $T_1$, varying the distance $L$ or the temperature $T_2$ of the hotter disc) collapse, if $\varepsilon$ is plotted versus the product $L T_2$. This experiment will be described in detail elsewhere; here we use it to estimate the relevant NF radiative heat transfer corrections. Fig.~\ref{NierFieldCu} allows to determine $\varepsilon$ and estimate the NF enhancement of heat flux radiated from the bottom plate for any relevant cryogenic RBC experiment. We note in passing that for flat cryogenic cells this simplified approach is sufficient. It is easy to show (see Fig.~\ref{NierFieldCu}) that the NF enhancement of the radiative heat transfer, for typical flat cryogenic cells used to determine the critical Rayleigh number, is only slightly larger than the standard FF radiation ($\varepsilon \simeq$ 0.2 \%), leading to negligibly small corrections for the data, such shown in Ref.~\cite{Lees} or in a number of experiments cited in Ref.~\cite{Metcalfe}. We therefore conclude that neither the FF nor the NF radiative heat transfer appreciably affects the total heat transfer in cryogenic helium RBC experiments.

\section{Conclusions}

In order to classify the role of thermal radiation in Rayleigh-B\'{e}nard convection, we have analyzed a few selected, well-documented RBC experiments, performed in cylindrical cells of various aspect ratios with gaseous working fluids at ambient and cryogenic temperatures. Utilizing the approach of Ref.~\cite{Hogg} for fully transparent working fluids, we show that in some turbulent high ${\rm{Ra}}$ RBC experiments the FF radiative heat transfer ought to be seriously taken into account, as it affects the Nusselt numbers inferred. Thus it appreciably changes the exponent $\gamma$ in the local scaling relation ${\rm{Nu}} \propto {\rm{Ra}}^\gamma$ in turbulent RBC experiments utilizing gaseous helium, nitrogen or sulphur hexafluoride near ambient temperatures as working fluids.

While for transparent working fluids evaluation of appropriate corrections is relatively straightforward, experiments utilizing gases absorbing thermal radiation as working fluids (e.g., sulphur hexafluoride) require special care. Thermal radiation as well as its possible absorption inside the RBC cell ought to be taken into account when planning accurate and reliable RBC experiments.

As for the cryogenic experiments performed at temperatures of a few K with gaseous or liquid helium as working fluids, we find that both FF and the enhanced NF radiative transport contribute to the total heat transport through the cell by a negligibly small amount (much smaller than uncertainties associated with typical corrections due to walls, plates etc.) and, for all cases discussed by us, the radiative heat transfer can be safely neglected.

\section*{Acknowledgements}
We thank M. Macek, M. La Mantia and K.R. Sreenivasan for stimulating discussions. The support of Czech Science Foundation under GA\v{C}R 17-03572S is acknowledged.

\section*{Appendix}

Following Ref.~\cite{Siegel} and for our particular case the Thesis of D. J. Hogg \cite{Hogg}, we give the derivation of the model used to calculate the part of heat flux carried from the heated bottom plate of the RBC cell by FF thermal radiation. The net energy gain/loss per unit area, the radiosity,  $J_{i}$, of a radiating, opaque, gray and diffuse surface $i$ held at temperature $T_{i}$ is a sum of two components: emitted intensity and reflected intensity. Total radiosity is then
\begin{equation}
\label{eq:radiosity_1}
J_{i} = e_i \sigma T_{i}^{4} + ( 1 - e_i ) H_{i}\,,
\end{equation}
\noindent where $e_i$ is the emissivity, $1-e_i$ expresses reflectivity, $\sigma$ is the Stefan-Boltzmann
constant and $H_{i}$ is the intensity of radiation incident on surface $i$.

\begin{figure}[th]
\centering
\includegraphics[width=.7\linewidth]{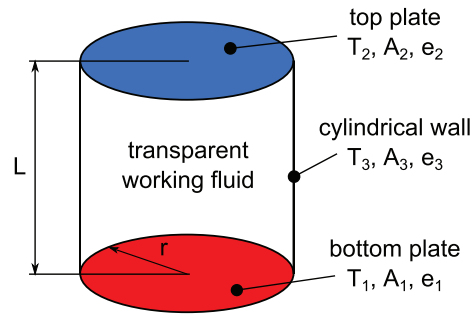}
\caption{The sketch of an experimental cell considered for calculation of thermal radiation in RBC experiments}.
\label{FigApp}
\end{figure}

The intensity of radiation incident on surface $i$, $H_{i}$, depends on the geometry of the entire considered system, which in this particular case is a simple cylindrical RBC cell, see Fig.~\ref{FigApp}. Generally, to characterize the geometry of the system, the view factors $F_{j \rightarrow i}$ are used; the total intensity of incident radiation upon surface $i$ is the sum of the intensities of energy from surface $j$ at surface $i$
\begin{equation}
\label{eq:sum_of_intensities}
H_{j \rightarrow i} = \frac{1}{A_{i}} \sum_{j=1}^{N} F_{j \rightarrow i} A_{j} J_{j}\,,
\end{equation}
\noindent where $A_{j}$ is the area of the surface $j$. Using the reciprocity theorem $A_{A} F_{A \rightarrow B} = A_{B} F_{B \rightarrow A}$ arising from the theory of view factors, we can simplify Eq.~(\ref{eq:sum_of_intensities}). After substitution to Eq.~(\ref{eq:radiosity_1}), we get
\begin{equation}
\label{eq:soustava_rovnic}
J_{i} = e_{i} \sigma T_{i}^{4} + ( 1 - e_{i} ) \sum_{j=1}^{N} F_{i \rightarrow j} J_{j}\,.
\end{equation}
The net heat transfer from surface $i$, $Q_{i}$, can then be calculated from
\begin{equation}
Q_{i} = A_{i} (J_{i} - H_{i} ) = \frac{A_{i} e_{i} }{1 - e_{i} } (\sigma T_{i}^{4} - J_{i} ).
\label{eq:Qi}
\end{equation}
In cylindrical RBC cells (see Fig.~\ref{FigApp}), three view factors are needed: disc to parallel disc $F_{1 \rightarrow 2}$, base of cylinder to inside of cylinder $F_{1 \rightarrow 3}$ and inside of cylinder to inside of cylinder $F_{3 \rightarrow 3}$ where index 1 is for the bottom plate, 2 is for the top one and 3 is for the inner surface of the sidewall.

The view factors for two parallel disks of equal radii $r$ separated by a distance $L$ can be written as
\begin{equation}
F_{1 \rightarrow 2} = F_{2 \rightarrow 1} = \frac{1}{2} \left[ X - (X^2 - 4)^{\frac{1}{2}} \right]\,,
\end{equation}
where $X = (2R^2+1)/R^2$ and $R = r/L$.

To describe the view factor $F_{1 \rightarrow 3}$ for the base of cylinder to inside of cylinder with radius $r$ and height $L$, another dimensionless quantity is defined as $H = L/2r$. With this, the view factor becomes
\begin{equation}
F_{1 \rightarrow 3} = 2H \left[ (1+H^{2})^\frac{1}{2} - H \right]\,.
\end{equation}
Next, the view factor from the inside of a cylinder to the inside of the same cylinder is

\begin{equation}
F_{3 \rightarrow 3} = (1+H) - \sqrt{1+H^2}
\end{equation}
This last view factor isn't zero because of the concavity of the cylinder. On the other hand, view factors $F_{1 \rightarrow 1}$ and $F_{2 \rightarrow 2}$ are zero.

The last view factor from inside of cylinder to base of the cylinder can be calculated with the rule saying that the sum of all view factors from one surface is equal to one. In that case, view factor $F_{3 \rightarrow 1}$ is
\begin{equation}
F_{3 \rightarrow 1} = \frac{1-F_{3 \rightarrow 3}}{2}\,.
\end{equation}
With known symmetries for view factors in a cylindrical cell, the system of linear equations generally expressed as Eq.~(\ref{eq:soustava_rovnic}) is simplified to the following three equations:
\begin{align}
J_{1} &= e_{1} \sigma_{1} T^{4}_{1} + (1 - e_{1})(F_{1 \rightarrow 2} J_{2} + F_{1 \rightarrow 3}J_{3}) \label{eq:lin1} \\
J_{2} &= e_{2} \sigma_{2} T^{4}_{2} + (1 - e_{2})(F_{2 \rightarrow 1} J_{1} + F_{2 \rightarrow 3}J_{3}) \label{eq:lin2} \\
J_{3} &= e_{3} \sigma_{3} T^{4}_{3} + (1 - e_{3})(F_{3 \rightarrow 1} J_{1} + F_{3 \rightarrow 2}J_{2} + F_{3 \rightarrow 3} J_{3})
\label{eq:lin3}
\end{align}
\noindent The $Q_1$ values given in Tab.~\ref{tab:examples} are evaluated using Eq.~(\ref{eq:Qi}) by substituting solutions of Eqs.~(\ref{eq:lin1})-(\ref{eq:lin3}).
\bibliographystyle{unsrt}
\bibliography{RadiationPRE}
\end{document}